\documentclass[aps,twocolumn,prb,amssymb,floatfix]{revtex4}

\usepackage {graphicx}
\usepackage{mathptmx}

\begin{document}
\title{A pulsating regime of magnetic deflagration}

\author {M. Modestov, V. Bychkov, M. Marklund}

\affiliation {Department of Physics, Ume{\aa} University, SE-901 87
Ume{\aa}, Sweden}

\begin{abstract}
The stability of a magnetic deflagration front in a collection of molecular
magnets, such as Mn$_{12}$-acetate, is considered. It is demonstrated that stationary
deflagration is unstable with respect to one-dimensional
perturbations if the energy barrier of the magnets is sufficiently
high in comparison with the release of Zeeman energy at the front;
their ratio may be interpreted as an analogue to the Zeldovich number, as found in problems of 
combustion. When the Zeldovich number exceeds a certain
critical value, a stationary deflagration front becomes unstable
and propagates in a pulsating regime. Analytical estimates for the
critical Zeldovich number are obtained. The linear  stage of the instability
 is investigated numerically by solving the eigenvalue
problem. The nonlinear stage is studied using direct numerical
simulations. The parameter domain required for experimental
observations of the pulsating regime is discussed.
\end{abstract}

\maketitle

\section{Introduction}

 Molecular magnets have been a subject of
intense experimental and theoretical studies  for almost two decades, see Refs.\
[\onlinecite{Gatteschi-review-03, Barco-review-05}] for a review. The field
belongs to mesoscopic physics as the typical length scales are between micro- and macroscopic. Such
an intermediate parameter regime provides unique conditions
for observing quantum and classical phenomena acting together.
Besides, a large magnetic spin of a single molecule makes molecular
magnets natural candidates for novel magnetic storage media and
quantum computing \cite{Leuenberger-01, Tejada-01}.

Two prominent and well investigated materials in this research field are
$\rm{Mn}_{12}$-acetate and $\rm{Fe}_{8}$, both which demonstrate
super-paramagnetic properties \cite{Gatteschi-review-03}. They
both have a large spin at the ground state ($S=10$) and strong
magnetic anisotropy with  10 pairs of degenerated levels corresponding to
positive and negative projections $S_{z} $ on a chosen axis, and
an additional level with $S_{z}=0$ \cite{Sessoli-Nature-93,
Paulsen-JMMM-95, Friedman-PRL-96, Thomas-Nature-96}.
 When an external magnetic field is applied
to a sample of molecular magnets along the crystal axis, then spin orientation in the direction of the
field becomes preferable. At low temperature all molecules
populate only one level (e.g. $S_{z}=10$) and the magnetization
reaches the saturation value. If the direction of the external
magnetic field changes to the opposite one, then the former ground
state of the molecules becomes metastable with an increased
potential energy (the Zeeman energy). An energy barrier hinders direct transition of the molecules
from the metastable state to the new ground state, which, therefore,
can occur as a 
thermal relaxation or as avalanches. Thermal relaxation goes slowly
and uniformly in space for the whole sample. At temperatures of a few Kelvin this process may be
neglected, since its characteristic
time is rather large, e.g., about two months for $\rm{Mn}_{12}$ at
2 K \cite{Sessoli-Nature-93}. However, many experimental works on
molecular magnets demonstrated quite fast transition in a form of
avalanche with the characteristic time about few milliseconds
\cite{Forminaya-97,Barco-99}. Recent detailed studies of the
avalanche revealed that the spin reversal does not happen
simultaneously within the whole sample in that case, but it occurs in a narrow
zone propagating as a front at a velocity of several meters per
second \cite{Suzuki-05, Hernandez-PRL-05, McHugh-07,
Garanin-Chudnovsky-2007, Hernandez-08, Villuendas-08, McHugh-09,Decelle-09}.
The avalanche is accompanied by significant heat release similar
to a deflagration wave in combustion and for this reason it was
named ``magnetic deflagration''. 

In slow combustion, the
deflagration front corresponds to a thin zone of chemical
reactions separating the cold fuel mixture and the hot burned products
\cite{Zeldovich et al-85}. The released energy is transmitted to
the cold fuel mixture due to thermal conduction; the temperature of
the  fuel mixture increases, which stimulates chemical
reactions. The slow combustion front propagates in gaseous
mixtures with a substantially subsonic velocity within the range
from several centimeters to several meters per second
\cite{Zeldovich et al-85}. Analogously, in magnetic deflagration, the energy of
chemical reactions is replaced by the Zeeman energy of the
molecular magnets in an external magnetic field. The released energy is distributed to the
neighboring layers by thermal conduction, which increases the temperature
of the originally cold medium and leads to much higher
probability of spin transition. In recent studies, Garanin and
Chudnovsky \cite{Garanin-Chudnovsky-2007} developed a theoretical
description of a planar stationary magnetic deflagration with
maximal possible release of the Zeeman energy (they called such a
regime ``full burning''). Since then, a number of papers have been
devoted to ignition techniques and speed measurements of the
magnetic deflagration in $\rm{Mn}_{12}$
\cite{McHugh-07,Hernandez-08, Villuendas-08, McHugh-09,
Decelle-09}. Magnetic avalanches have been also observed in other
materials, like the intermetallic compound $\rm{Gd}_{5}\rm{Ge}_{4}$
\cite{Velez-10}. However, the magnetic ``burning'' does not
necessarily have to be complete. As we show in the present paper,
the Zeeman energy release depends on the external magnetic field
and on the initial concentration of active molecules. A
sufficiently low energy release may lead to a new pulsating regime
of magnetic deflagration.

The aim of the present work is to study the stability of a planar
stationary front of magnetic deflagration. Here, we demonstrate
that stationary deflagration is unstable with respect to
one-dimensional perturbations if the Zeeman energy release at the
front is sufficiently low in comparison with the energy barrier of
the spin transition. The condition of the instability may be
formulated using the Zeldovich number, similar to the case of combustion of
solid propellants \cite{Shkadinsky-71, Matkovsky-78, Zeldovich et
al-85, Frankel-94, Bychkov-Liberman-94, Bychkov-review}. In the
problem of magnetic deflagration, the Zeldovich number is essentially given by the 
ratio of the energy barrier of the molecular magnets and the 
temperature at the deflagration as determined by the Zeeman energy
release. When the Zeldovich number exceeds a certain critical value, a
stationary deflagration front becomes unstable and propagates in a
pulsating regime. We obtain analytical estimates for the critical
Zeldovich number. We investigate the linear stage of the
instability numerically by solving the  eigenvalue problem. We study
the nonlinear stage using direct
numerical simulations. We also discuss the experimental parameters
required for observations of the pulsating regime.

\section{A planar stationary front with incomplete Zeeman energy release}

We consider a system of molecular magnets
placed in an external magnetic field taking $\rm{Mn}_{12}$-acetate as a particular example.
The simplified Hamiltonian
of the system has been suggested to take the form
\cite{Garanin-Chudnovsky-2007}
\begin{equation}
\label{eq1}
\mathcal{H} = - DS_{z}^{2} - g\mu_{B}H_{z}S_{z},
\end{equation}
\noindent where $S$ denotes the spin, $D\approx0.65K$ is the magnetic
anisotropy constant, $g\approx1.94$ is the gyromagnetic factor,
$\mu_{B}$ is the Bohr magneton, and $H_{z}$ is the external magnetic
field.
\begin{figure}
\includegraphics[width=3.4in,height=2.4in]{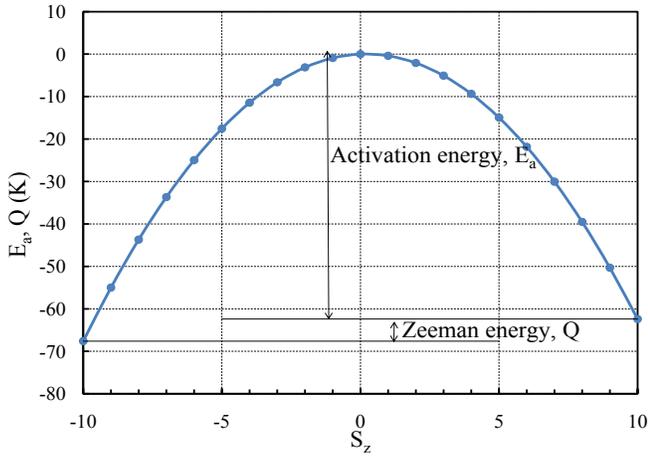}
\caption{The energy levels for the magnetic molecule system of
$\rm{Mn}_{12}$ in the external magnetic field $H_{z}=0.2T$}
\end{figure}
The energy levels for a molecule magnet
$\rm{Mn}_{12}$ in the external field of $H_{z}=0.2T$ are depicted in
Fig. 1. In Fig. 1 we indicate the Zeeman energy $Q$ and the energy
barrier $E_{a} $ of the transition from the metastable state to the
ground state, which plays the role of the activation energy for magnetic
deflagration (both values are presented in temperature units). Using
the Hamiltonian Eq. (\ref{eq1}) we find how these two energies
depend on the magnetic field, according to
\begin{equation}
\label{eq2}
 E_{a} = DS_{10}^{2} - g\mu _{B} H_{z} S_{10} + \frac{g^{2}}{4D}\mu
_{B}^{2} H_{z}^{2}
\end{equation}
and 
\begin{equation}
\label{eq3}
 Q = 2g\mu_{B} H_{z} S_{z}.
\end{equation}
The last term in Eq. (\ref{eq2}) is quite small as
compared to the first two terms, so the activation energy decreases
with increase of the magnetic field. However, the Zeeman
energy increases linearly with the magnetic field. If the magnetic
field is high enough, the energy difference between the ground
state ($S_{z}=-10)$ and the metastable state $(S_{z}=10)$ is
rather large, so that all the molecules tend to occupy the lowest
energy level, i.e. $S_{z}=-10$ in Fig. 1. When the direction of the
external magnetic field switches to the opposite one, then the
ground state and the metastable state exchange places and all the
molecules tend to change their spin projection to the opposite
one. Garanin and Chudnovsky identified such a transition as ``full
burning'' of molecular magnets \cite{Garanin-Chudnovsky-2007}.
They also indicated that the regime of full burning is possible if
the magnetic field is stronger than a certain critical value.
Still, the external magnetic field is a controlled parameter of
the experiments, which allows the possibility of incomplete
burning in magnetic deflagration. We also point out that the
analytical method used to calculate the magnetic deflagration
speed in [\onlinecite{Garanin-Chudnovsky-2007}] is rigorous only in the
limit of an activation energy large compared to the energy
release, a condition which is satisfied to a much better degree in the regime of
incomplete burning. In the present paper we are interested in the
regime of incomplete burning in magnetic deflagration achieved for
a sufficiently low external magnetic field. In that case the
Zeeman energy is rather low, so that we obtain the fraction
of molecules on the first energy level according to 
\begin{equation}
\label{eq4} n = \frac{1}{\exp \left( {Q / T} \right) + 1},
\end{equation}
where total number of molecules in all energy levels
corresponds to unity. Still, in the typical experimental
conditions, population of levels above the first one is
negligible. At the same time, the fraction at the first level, Eq.
(\ref{eq4}), may be significant in a low magnetic field and cannot
be neglected. As we will see, the stability of magnetic deflagration
is sensitive to the activation energy scaled by the energy release
in the process. This ratio increases also when the active
$\rm{Mn}_{12}$ molecules are ``diluted'' with some neutral media
as it was performed experimentally in, e.g., [\onlinecite{Artus, Schlegel,
Makarova}]. When mixed carefully with other compounds, the
$\rm{Mn}_{12}$ molecules retain their magnetic properties. Since
the heat release in magnetic deflagration happens only due to
active magnet molecules, then the neutral compound reduces total
energy release and the final temperature of the sample, $T_{f} $,
thus increasing the scaled activation energy $E_{a}/T_{f} $. As a
result, the scaled activation energy in magnetic deflagration
becomes a free parameter, which may be controlled in the
experiments by the external magnetic field and by the level of
dilution.

The governing equations for magnetic deflagration are
\cite{Garanin-Chudnovsky-2007}
\begin{equation}
\label{eq5}
\frac{\partial E}{\partial t} = \nabla \cdot \left(
{\kappa \nabla E} \right) - f Q\frac{\partial n}{\partial t} ,
\end{equation}
\begin{equation}
\label{eq6} \frac{\partial n}{\partial t} = - \frac{1}{\tau _{R}}
\exp \left( { - \frac{E_{a}} {T}} \right) \left[ n - \frac{1}{\exp
\left( {Q / T} \right) + 1} \right]
\end{equation}
\noindent where $E$ is the thermal phonon energy, $\kappa $ is
thermal diffusion of energy, $Q$ is the Zeeman energy release in
temperature units, $f \le 1$ indicates possible reduction of the
energy release because of dilution of active molecules ($f = 1$
corresponds to zero dilution), $n$ is fraction of magnetic molecules
in the metastable state, $E_{a} $ is the energy barrier of tunneling
measured in temperature units and $\tau _{R} $ is a constant of time
dimension. In magnetic deflagration, thermal diffusion and heat
capacity depend strongly on temperature. Following
\cite{Garanin-Chudnovsky-2007}, we take the heat capacity in the
classical form corresponding to phonons \cite{Kittel}
\begin{equation}
\label{eq7} C = Ak_{B} \left( \frac{T}{\Theta _{D}}\right)^{\alpha},
\end{equation}
\noindent where $\Theta _{D} $ is the Debye temperature, with
$\Theta_{D}=38K$ for $\rm{Mn}_{12}$, $k_{B}$ is the Boltzmann
constant, $A=12\pi^{4}/5$ corresponds to the simple crystal model,
$\alpha $ is the problem dimension (we take $\alpha=3$ in most of
the calculations, which corresponds to the 3D geometry).
Dependence of thermal diffusion on temperature may be taken in the
form $\kappa \propto T^{-\beta}$, where parameter $\beta$ was
considered in Refs. [\onlinecite{Garanin-Chudnovsky-2007, Garanin-92}]
within the range between 0 and 13/3. In the present paper we take
$\beta $ within the same range, and show that it has minor
influence on the deflagration stability. Using the definition of
heat capacity, $C=dE/dT$, we find the phonon energy as a function
of temperature
\begin{equation}
\label{eq8} E = \frac{Ak_{B}} {\alpha + 1}\left( \frac{T}{\Theta
_{D}} \right)^{\alpha} T.
\end{equation}
An important parameter of deflagration dynamics is determined by
the ratio of the energy barrier (in temperature units) and the
temperature in the hot region, $E_{a} / T_{f} $. A combustion
counterpart of this value is related to the activation energy of
chemical reaction, which is typically rather large. In the case of
complete burning in magnetic deflagration of $\rm{Mn}_{12} $ this
parameter was evaluated in [\onlinecite{Garanin-Chudnovsky-2007}] as
$E_{a} / T_{f} \approx 6$, which is not a large value. At the same
time, this parameter may increase almost without limits at low
magnetic fields and for considerable dilution of active molecular
magnets.
When the ratio $E_{a} / T_{f} $ is very large, the
transition from the metastable to stable states goes in a thin
region where the Arrhenius function in Eq. (\ref{eq6}) is
different from zero. In this limit we may obtain an analytical
solution to Eqs. (\ref{eq5}), (\ref{eq6}) by the classical method
of Zeldovich-Frank-Kamenetsky \cite{Zeldovich et al-85}, which was
applied to magnetic deflagration in
[\onlinecite{Garanin-Chudnovsky-2007}]. In a general case of finite $E_{a}
/ T_{f} $, the system Eqs. (\ref{eq6}), (\ref{eq6}) may be solved
numerically as an eigenvalue problem.

First, we calculate the final temperature in the hot region behind the
magnetic deflagration front. Using the energy conservation and
Eqs. (\ref{eq4}), (\ref{eq5}), (\ref{eq8}) we obtain
\begin{eqnarray}
 \frac{AT_{0}^{\alpha + 1}}{fQ(\alpha +
1)\Theta_{D}^{\alpha}} + 1 - \frac{1}{\exp\left( Q / T_{0} \right) +
1} = \nonumber \\ \frac{AT_{f}^{\alpha + 1}}{fQ(\alpha +
1)\Theta_{D}^{\alpha}} + \frac{1}{\exp \left( {Q / T_{f}} \right) +
1}\label{eq9}.
\end{eqnarray}
Equation (\ref{eq9}) determines the final temperature $T_{f} $ as
a function of the magnetic field, the dilution factor $f$ and other
parameters of the process. The first terms on both sides stand for
the thermal energy, though initial thermal energy is usually
rather small. The other terms indicate the fractions of molecules
occupying the first energy level. In the case of full burning considered in [\onlinecite{Garanin-Chudnovsky-2007}] this
equation becomes much simpler and may be solved analytically. In
the case of incomplete burning, a numerical solution to Eq.
(\ref{eq9}) is required. In Fig. 2 we show the numerical solution
to Eq. (\ref{eq9}), i.e. the final temperature and the scaled
activation energy versus the magnetic field for two dilution
factors $f = 1;\,1 /3$.
\begin{figure}
\includegraphics[width=3.4in,height=2.6in]{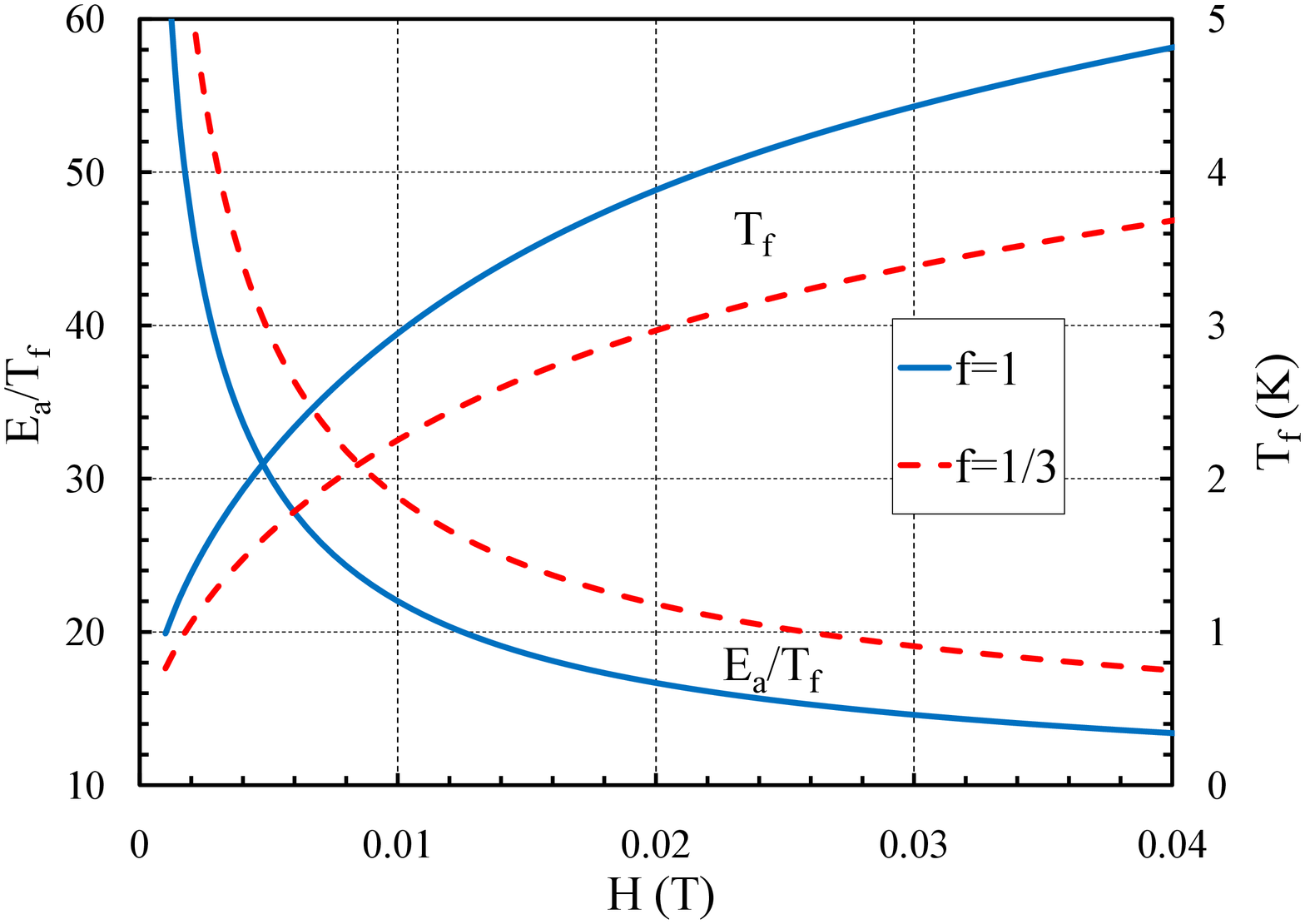}
\caption{The final temperature and the scaled activation energy in
magnetic deflagration versus the magnetic field for the initial
temperature $T_{0} = 0.2K$ and two dilution factors $f = 1;\,1 /
3$.}
\end{figure}
In Fig. 2 we take the initial temperature $T_{0} = 0.2K$, though it
has minor influence on the result. The dilution factor $f=1$ stands
for pure $\rm{Mn}_{12}$ media, while $f=1/3$ means that the average
energy release is 3 times lower in comparison with the pure
substance. The final temperature increases with the magnetic field,
while the scaled activation energy decreases. The change in the
scaled activation energy due to the dilution factor may be also
quite strong.

We consider a stationary solution to Eqs. (\ref{eq5}), (\ref{eq6})
in the form of a planar front propagating with velocity $U_{f}$. To
be particular, we assume that the front moves along the x-axis in
the negative direction as shown in Fig. 3. Taking the reference
frame of the front, we find
\begin{equation}
\label{eq10} U_{f} \frac{d}{dx}\left( {E + fQn} \right) =
\frac{d}{dx}\left( {\kappa \frac{dE}{dx}} \right),
\end{equation}
\begin{equation}
\label{eq11} U_{f} \frac{dn}{dx} = - \frac{n}{\tau _{R}} \exp \left(
{- \frac{E_{a}} {T}} \right){\left[ {n - \frac{1}{\exp \left( {Q /
T} \right) + 1}} \right]}.
\end{equation}
\noindent
Integrating Eq. (\ref{eq10}) from the initial to final
states (labels 0 and f, respectively) and neglected the initial thermal energy, we obtain the final energy as
\begin{equation}
\label{eq12} E_{f} = fQ\left( {n_{0} - n_{f}}  \right).
\end{equation}
\noindent
Integral to Eq. (\ref{eq10})
specifies the internal structure of the transition zone
\begin{equation}
\label{eq13} U_{f} fQ\left( {n_{0} - n_{f}}  \right) = \kappa _{f}
{\frac{dE}{dx}}.
\end{equation}
\noindent In the limit of a large activation energy, $E_{a} / T_{f} \gg 1$, the transition region may be presented as a surface of weak
discontinuity, where energy and temperature are continuous and tend
to their maximal values $E \to E_{f} $, $T \to T_{f} (E_{f} )$,
while their derivatives experience jump. In this limit one obtains
the magnetic deflagration velocity \cite{Zeldovich et al-85}
\begin{equation}
\label{eq14} U_{f} = \sqrt {\frac{\kappa _{f}} {Z\tau_{R}}} \exp
\left( { - \frac{E_{a}} {2T_{f}} } \right),
\end{equation}
\noindent where
\begin{equation}
\label{eq15} Z = \frac{E_{a}} {T_{f}} \frac{fQ\left( {n_{f} - n_{0}}
\right)} {C_{f} T_{f}} = \frac{1}{\left( {\alpha + 1} \right)}
\frac{E_{a}} {T_{f}}
\end{equation}
\noindent plays the role of the Zeldovich number; the last
relation in Eq. (\ref{eq15}) follows from Eqs. (\ref{eq7}),
(\ref{eq8}) and (\ref{eq12}). Equation (\ref{eq14}) for the
deflagration velocity coincides with Eq. (109) in Ref.
[\onlinecite{Garanin-Chudnovsky-2007}]. With the accuracy of a factor of
$1 / \left( {\alpha + 1} \right)$, the Zeldovich number shows the
ratio of the activation energy and the final temperature in the
deflagration front. Strictly speaking, the approach of a thin
transition zone holds as long as Zeldovich number is large $Z \gg
1$, which is not applicable to the regime of full burning. We can
find the dependence of the Zeldovich number upon the main
experimental parameters of the problem taking into account Eqs.
(\ref{eq2}),(\ref{eq3}),(\ref{eq7}),(\ref{eq8}) and (\ref{eq12})
\begin{eqnarray}
 Z = \left[ \frac{A}{2(\alpha + 1)^{\alpha + 2}\Theta
_{D}^{\alpha} g\mu_{B} S_{z} H_{z} f\left(n_{0} - n_{f} \right)}
\right]^\frac{1}{\alpha + 1} \times \nonumber \\ \left( DS_{10}^{2}
- g\mu_{B} H_{z} S_{10} + \frac{g^{2}}{4D}\mu_{B}^{2} H_{z}^{2}
\right)\label{eq16}.
\end{eqnarray}
\noindent Equation (\ref{eq16}) shows how the Zeldovich number
depends on the magnetic field and the dilution factor. In particular, high values of the Zeldovich number
correspond to the low field strength. For a simplified quantitative
estimate one can neglect the magnetic terms in the second couple of
parentheses, which provide a contribution less than few percents for
fields about 0.1 T and smaller, and find an evaluation for the
Zeldovich number
\begin{equation}
\label{eq17} Z \approx 1.3{\left[ {H_{z} f\left( {n_{0} - n_{f}}
\right)} \right]}^{ - 1 / 4},
\end{equation}
\noindent where $H_z$ is the magnetic field in tesla.

\section{Analytical estimates for the pulsation instability}

In this section we obtain an analytical scaling for the 1D
instability of the magnetic deflagration front in the model of a
thin transition zone, i.e. at a large Zeldovich number $Z \gg 1$. In
that case, according to Eq. (\ref{eq14}), the deflagration velocity
is extremely sensitive to temperature variations in the transition
zone. In comparison with the Arrhenius function $\exp \left( { -
E_{a} / 2T_{f}} \right)$, all other parameters in Eq. (\ref{eq14})
may be treated as constant. Temperature in the transition zone may
vary because of the front perturbations, and the instant front
velocity may be written as
\begin{equation}
\label{eq18} U_{t} = U_{f} \exp \left( \frac{E_{a}}{2T_{f}}  -
\frac{E_{a} }{2T_{t}} \right),
\end{equation}
where label $f$ refers to the stationary case, whilst
$t$ indicates the time-dependent temperature in the infinitely
thin transition zone. Position of the transition zone, $x = \phi (t)$, in the
stationary reference frame  is determined by the
equation
\begin{equation}
\label{eq19} \frac{\partial \phi} {\partial t} = - U_{f} {\left[
\exp \left( \frac{E_{a}} {2T_{f}} - \frac{E_{a}} {2T_{t}} \right) -
1 \right]},
\end{equation}
where the first minus sign indicates propagation of the front
in the negative direction. Boundary conditions at the transition
region may be found similar to \cite{Zeldovich et al-85,
Matkovsky-78}. Integrating Eq. (\ref{eq5}) twice over the
transitional zone, we obtain continuous energy and temperature
\begin{equation}
\label{eq20} E_{\phi +}  = E_{\phi -}  ,
\end{equation}
where the labels $\phi _{\pm}  $ correspond to hot and
cold sides of the transition zone. The first derivative of energy
(i.e. energy flux) experiences jump at the interface as
\begin{eqnarray}
\label{eq21}
 - U_{f} Qn_{0} \exp \left( \frac{E_{a}} {2T_{f}} - \frac{E_{a}
}{2T_{t}} \right) = \nonumber \\ \kappa _{f} \left( \frac{\partial E}{\partial x}
\right)_{\phi +}  - \kappa _{f} \left( \frac{\partial E}{\partial x}
\right)_{\phi -} .
\end{eqnarray}
We investigate the linear stability of the stationary
deflagration and consider small perturbations of all variables in
the exponential form $\tilde {E} \propto \exp (\sigma t)$, where the
growth rate $\sigma $ may have both real and imaginary parts. We
solve the stability problem analytically in the limit of a thin
transition zone similar to [\onlinecite{Zeldovich et al-85}]. Outside the
transition zone the perturbed equations (\ref{eq5})-(\ref{eq6}) are
\begin{equation}
\label{eq22} \sigma \tilde {E} + U_{f} \frac{\partial \tilde
{E}}{\partial x} = \frac{\partial ^{2}}{\partial x^{2}}\left(
{\kappa \tilde {E}} \right),
\end{equation}
\begin{equation}
\label{eq23} \sigma \tilde {n} + U_{f} \frac{\partial \tilde
{n}}{\partial x} = 0.
\end{equation}
\noindent Behind the transition zone we have $\tilde {n} = 0$.
Similar to the respective combustion problem \cite{Zeldovich et
al-85}, we consider first a hypothetic case of constant coefficient
of thermal conduction $\kappa = const = \kappa _{f} $. Solving Eq.
(\ref{eq22}) outside the transition zone with $\tilde {E} \propto
\exp (\mu x)$ we find two modes
\begin{equation}
\label{eq24} \mu ^{2} - \frac{U_{f}} {\kappa _{f}}\mu -
\frac{\sigma} {\kappa _{f}} = 0,
\end{equation}
\begin{equation}
\label{eq25} \mu _{0,1} = \frac{U_{f}} {2\kappa _{f}}\pm \sqrt {
\frac{U_{f}^{2} }{4\kappa_{f}^{2}} + \frac{\sigma} {\kappa_{f}} } ,
\end{equation}
\noindent with labels 0 and 1 indicating media ahead and behind
the transition zone. In the case of magnetic deflagration,
thermal conduction decreases strongly with temperature, $\kappa
\propto T^{ - \beta} $. Still, the mode behind the transition zone
with $\mu _{1} < 0$ is an exact solution to the linearized Eq.
(\ref{eq19}) even in that case, since temperature behind the
transition zone is uniform in the stationary deflagration. On the
contrary, the mode ahead of the transition zone with $\mu _{0} >
0$ is only an approximation. Similar to the combustion theory
\cite{Zeldovich et al-85}, we may define the parameter $L_{f}
\equiv \kappa _{f} / U_{f} $ as the deflagration front thickness
related to thermal conduction in the hot region. At the same time,
the characteristic length scale of temperature profile in the
stationary deflagration increases in the cold layers as $\kappa
(T) / U_{f} $, which allows to consider them as quasi-uniform with
respect to the perturbations. Indeed, the mode ahead of the
transition zone describes decay of perturbations at the length
scale $\mu_{0}^{-1}$ with
\begin{equation}
\label{eq26} \mu _{0} = \frac{1}{2L_{f}} {\left[ {1 + \sqrt {1 +
4\sigma L_{f} / U_{f}}} \right]} \propto \frac{1}{L_{f} }.
\end{equation}
\noindent
Therefore, investigating the 1D instability analytically, it is
justified to consider only the heating layer of the size about
$L_{f} $ close to the transition zone and to treat the coefficient
of thermal conduction as approximately constant. The numerical
solution to the problem obtained below supports the analytical
approximation.

We consider perturbations of the boundary conditions (\ref{eq19}) --
(\ref{eq21}) at the transition surface
\begin{equation}
\label{eq27} \sigma \phi = - U_{f} \frac{E_{a}} {2C_{f} T_{f}^{2}}
\tilde {E}_{\phi +}  ,
\end{equation}
\begin{equation}
\label{eq28} \tilde {E}_{\phi +}  = \tilde {E}_{\phi -}  + \phi
\left( {\frac{dE}{dx}} \right)_{\phi -}  = \tilde {E}_{\phi -} +
\phi n_{0} Q / L_{f} .
\end{equation}
\begin{eqnarray}
 - U_{f} Qn_{0} \frac{E_{a}} {2C_{f} T_{f}^{2}} \tilde {E}_{\phi +}  =
\kappa_{f} \left( \frac{\partial \tilde {E}}{\partial x}
\right)_{\phi +}  - \kappa_{f} \left( \frac{\partial \tilde
{E}}{\partial x} \right)_{\phi -} \nonumber \\ - \kappa_{f} \phi
\left( \frac{\partial ^{2}E}{\partial x^{2}} \right)_{\phi -} \quad
\quad \quad \quad \quad \label{eq29}.
\end{eqnarray}
\noindent The last equation may be also modified as
\begin{equation}
\label{eq30}
 - Qn_{0} {\frac{{E_{a}}} {{2C_{f} T_{f}^{2}}} }\tilde {E}_{\phi +}  = \mu
_{1} L_{f} \tilde {E}_{\phi +}  - \mu _{0} L_{f} \tilde {E}_{\phi -}
- \phi n_{0} \frac{Q} {L_{f}}
\end{equation}
\noindent taking into account the perturbation modes. Thus, we have
to solve the algebraic system of three equations (\ref{eq27}),
(\ref{eq28}) and (\ref{eq30}), taking into account the modes
(\ref{eq25}). After heavy but straightforward algebra we end up with
a simple quadratic equation
\begin{equation}
\label{eq31} \left( \frac{4\sigma L}{U_{f}} \right)^{2} -
\frac{4\sigma L}{U_{f}} \left(Z^{2} / 4 - 2Z - 1 \right) + 2Z = 0,
\end{equation}
\noindent which is similar to the respective result in the
combustion theory \cite{Zeldovich et al-85}. Equation (\ref{eq31})
describes the instability growth rate as a function of the
Zeldovich number. According to Eq. (\ref{eq31}), the instability
develops for $Z > 4 + 2\sqrt {5} \approx 8.5$. Close to this
critical value the real part of the growth rate $\rm{Re}\sigma$
goes to zero, while the imaginary part remains finite, $\omega =
\rm{Im}\sigma \ne 0$, which indicates the pulsation regime of the
instability. The obtained critical value of the Zeldovich number
corresponds to rather high scaled activation energy, $E_{a} /
T_{f} \approx 34$, see Eq. (\ref{eq15}), in comparison with the
values typical for the regime of full burning Ref.
[\onlinecite{Garanin-Chudnovsky-2007}]. Still, this value is attainable
experimentally for smaller magnetic fields and some dilution of
the active molecules. At even larger value of the Zeldovich
number, $Z = 11.7$, corresponding to $E_{a} / T_{f} = 46.8$, Eq.
(\ref{eq31}) demonstrates bifurcation, so that the instability
growth rate becomes purely real with zero imaginary part for $Z >
11.7$. Below we will compare the analytical scaling Eq.
(\ref{eq31}) to the numerical solution to the problem taking into
account finite width of the transition zone.

\section{Numerical solution to the stability problem taking into account
finite width of the transition zone}

In this section we solve the stability problem numerically taking into account finite width
of the transition zone. We introduce dimensionless variables and parameters
$$ \theta = T / T_{f} , \quad a = n / n_{0} , \quad \xi = x / L_{f}
,$$
\begin{equation}
\label{eq32}
\Theta = E_{a} / T_{f} , \quad \Delta = Q / T_{f} ,
\quad \kappa = \kappa _{f} \theta ^{ - \beta}
\end{equation}

\noindent and rewrite the evolution equations
(\ref{eq5})-(\ref{eq6}) to
\begin{eqnarray}
 \theta^{\alpha} \frac{\partial \theta} {\partial \tau}
 + \theta^{\alpha} \frac{\partial \theta} {\partial \xi} =
\frac{\partial}{\partial \xi} \left( \theta^{\alpha - \beta}
\frac{\partial \theta}{\partial \xi} \right) + \quad \quad \nonumber
\\ J\Lambda \exp \left(
 - \frac{\Theta}{\theta} \right) \left[ {a - \frac {1} {\exp
\left( \Delta / \theta \right) + 1 }} \right], \label{eq33}
\end{eqnarray}
\begin{equation}
\label{eq34} \frac{\partial a}{\partial \tau} + \frac{\partial
a}{\partial \xi } = - \Lambda \exp \left( - \frac{\Theta}{\theta}
\right) \left[ a - \frac{1}{\exp \left( \Delta / \theta \right) + 1
} \right],
\end{equation}
\noindent where $\Lambda = L_{f} / \left( {\tau_{R} U_{f}} \right)$
is an eigenvalue of the stationary problem and the designation $J$
corresponds to the ratio of Zeeman and thermal energies,
\begin{equation}
\label{eq35} J = \frac{fQ\Theta_{D}^{\alpha}} {Ak_{B} T_{f}^{\alpha
+ 1}} .
\end{equation}
\noindent It can be shown that the parameter $J$ is almost a
constant, $J \approx 1 / \left( {\alpha + 1} \right)$. We also
introduce the thermal flux, $\psi = \theta ^{\alpha - \beta}
\partial \theta / \partial \xi $, so that the stationary
deflagration is described by the system of equations
\begin{eqnarray}
 \frac{\partial \psi} {\partial \xi} = \theta^{\beta} \psi -
J\Lambda a\exp \left(-\frac{\Theta} {\theta} \right), \nonumber \\
 \frac{\partial \theta} {\partial \xi} = \theta^{\beta - \alpha
}\psi , \quad \quad \quad \quad \quad \quad \quad \label{eq36} \\
 \frac{\partial a}{\partial \xi} = - \Lambda a\exp \left(-
\frac{\Theta} {\theta}\right). \quad \quad \quad \nonumber
\end{eqnarray}
\noindent Typical profiles of scaled concentration, temperature and
energy release
\begin{equation}
\label{eq37}
 W = \Lambda \exp \left( - \frac{\Theta} {\theta} \right) \left[a
- \left( \exp \left( \frac{\Delta} {\theta} \right) + 1 \right)^{-1}
\right]
\end{equation}
\noindent are depicted in Fig. 3 for the magnetic deflagration
propagating to the left with $\beta = 3$. In Fig. 3, the scaled
activation energy is taken rather high, $\Theta = 30$, with the
initial temperature $\theta _{0} = 0.2$.
\begin{figure}
\includegraphics[width=3.4in,height=2.5in]{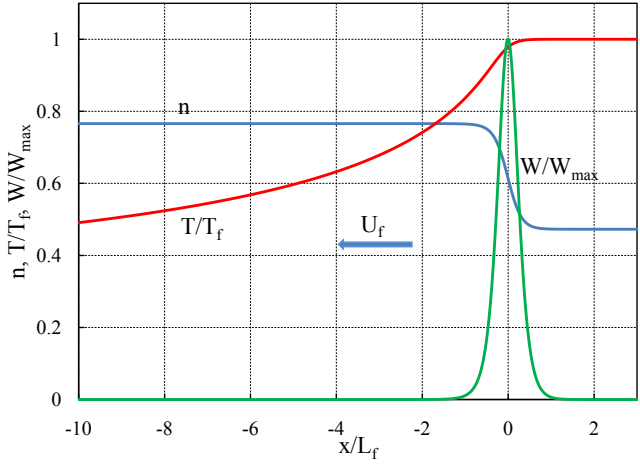}
\caption{Stationary profiles of concentration, temperature and
energy release. The plot parameters are $E_{a} / T_{f} = 30$, $\beta
= 3$, $\theta _{0} = 0.2$}
\end{figure}
The front velocity is determined by the eigenvalue of the problem
$\Lambda$ in Eq. (\ref{eq36}), which was computed numerically using
the shooting method similar to \cite{Modestov-ADLI-09}. We can see
in Fig. 3 that final number of molecules in the metastable state
behind the deflagration front is different from zero $n_{f} = 0.47$,
which indicates that burning is incomplete. Initial number of
molecules in the metastable state is different from unity too,
$n_{0} = 0.77$, due to the non-zero temperature ahead of the front
before switching of the magnetic field, see Eq. (\ref{eq4}). Figure
3 shows also that the total thickness of the deflagration front is
much larger than $L_{f}$. This difference in the characteristic
length scales should be attributed, first of all, to the temperature
dependence of thermal conduction. Still, even in the case of
constant thermal conduction typically used in combustion problems,
the effective flame thickness is almost order-of-magnitude larger
than the conventional definition for $L_{f} $, e.g. see
\cite{Akkerman.et.al.-2006}. The asymptotic temperature behavior
ahead of the front is described by a single differential equation,
derived from Eq. (\ref{eq33}), assuming that $\exp \left(-\Theta /
\theta \right)<<1$ in the cold matter
\begin{equation}
\label{eq38} \frac{\partial \theta} {\partial \xi} + \frac{\theta
^{\beta - \alpha}} {\alpha + 1} \left( \theta _{0}^{\alpha + 1} -
\theta ^{\alpha + 1} \right) = 0.
\end{equation}
\noindent Equation (\ref{eq38}) determines variations of the length
scale in the temperature profile.

Now we consider stability of the magnetic deflagration. We
investigate dynamics of small perturbations of the stationary
solution, so that all variables may be presented as $\varphi =
\varphi _{0} + \tilde {\varphi }\exp \left( {\gamma \tau}  \right)$.
Then the perturbed system (\ref{eq36}) is
\begin{eqnarray}
 {\frac{{\partial \tilde {\psi}} }{{\partial \xi}} } = \theta ^{\beta
}\tilde {\psi}  + \left( {\theta ^{\alpha} \sigma + \beta \theta ^{\beta -
1}\psi - JG} \right)\tilde {\theta}  - J\Lambda \exp \left( { -
{\frac{{\Theta}} {{\theta}} }} \right)\tilde {a}, \nonumber \\
 {\frac{{\partial \tilde {\theta}} }{{\partial \xi}} } = \theta ^{\beta -
\alpha} \tilde {\psi}  + \left( {\beta - \alpha}  \right)\theta
^{\beta - \alpha - 1}\psi \tilde {\theta},  \quad \quad \quad \quad
\quad \quad \quad \quad \quad \quad \nonumber
 \\
 {\frac{{\partial \tilde {a}}}{{\partial \xi}} } = - G\tilde {\theta}  -
\left( {\sigma + \Lambda \exp \left( { - {\frac{{\Theta}} {{\theta}}
}} \right)} \right)\tilde {a}, \;\; \quad \quad \quad \quad \quad
\quad \quad \quad
\label{eq39}
\end{eqnarray}
\noindent where the designation
\begin{eqnarray}
 G = \Lambda \exp \left(-\frac{\Theta}{\theta} \right)\frac{1}{\theta ^{2}}
\left[ \Theta a - \right. \quad \quad  \quad \quad \quad \quad \quad \quad \quad \quad \nonumber \\
\left. \left( {\Theta - \frac{\Delta} {\exp \left( {\Delta / \theta}
\right) + 1}} \right) \frac{1}{\exp \left( {\Delta / \theta} \right)
+ 1} \right] \nonumber
  \label{eq40}
\end{eqnarray}
\noindent has been introduced for brevity. Parameter $\gamma $
stands for the scaled instability growth rate, $\gamma = \sigma
L_{f} / U_{f} $. In order to specify boundary conditions for the
numerical solution, we find the decaying perturbation modes in the
uniform regions, $\tilde {\varphi}  \propto \tilde {\varphi} \exp
(\mu \xi )$. The system (\ref{eq39}) involves three modes in the
uniform regions: one in the hot matter with $\mu < 0$, $ \xi
\to + \infty$, and two in the cold matter with $\mu > 0$, $\xi \to - \infty $.
We integrate the system (\ref{eq39})
numerically three times corresponding to the three perturbation
modes: two times from the left (cold) side and one from the right
(hot) side. At some point, e.g., at the maximum of the energy
release, the solutions form a matrix, with the determinant
depending on the instability growth rate, $\gamma $. The condition
of zero matrix determinant specifies the growth rate $\gamma $ as
an eigenvalue of the system (\ref{eq39}). This method has been
 applied successfully in  studies of different hydrodynamic
instabilities, e.g. the Rayleigh-Taylor instability and the
Darrieus-Landau instability of combustion and laser ablation, as
well as for other plasma instabilities \cite{Travnikov-99,
Modestov-QMPRTI-09, Modestov-ADLI-09, Bychkov-MHDI-10}.

\section{Results and discussions}

Numerical solution to the stability problem is shown in Fig. 4 for
different values of the scaled activation energy proportional to
Zeldovich number as $E_{a} / T_{f} = 4Z$ for a 3D problem; other
deflagration parameters are the same as used in Fig. 3. As we can
see in Fig. 4, planar stationary magnetic deflagration is unstable
at sufficiently large values of the scaled activation energy; the
stability limit was calculated as $E_{a} / T_{f} = 28.2$.
\begin{figure}
\includegraphics[width=3.4in,height=2.6in]{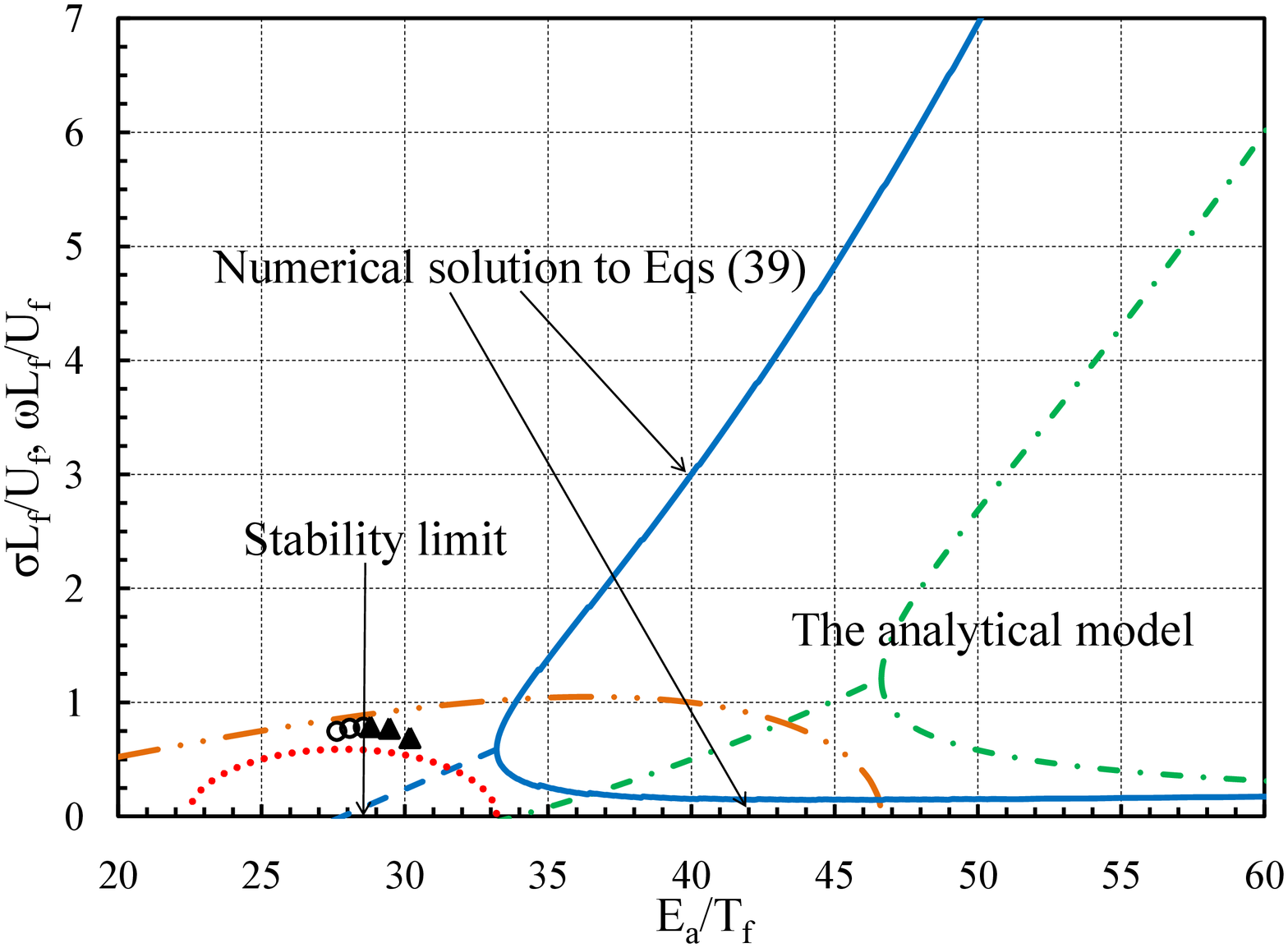}
\caption{The instability growth rate $\sigma $ and the perturbation
frequency $\omega $ versus the scaled activation energy calculated
for the same parameters as in Fig. 3. Solid lines show the domain of
zero frequency; the dashed and dotted lines correspond to the
regimes when perturbations have both real (growth rate, $\sigma )$
and imaginary part (frequency, $\omega )$ in the problem eigenvalue.
The dash-dotted lines present the analytical model Eq. (\ref{eq28}).
The markers show results of the direct numerical simulations: empty
circles stand for stable region and filled triangles represent the
unstable pulsating regime of the deflagration.}
\end{figure}
The instability domain consists of two parts separated by the
bifurcation point at $E_{a} / T_{f} = 33.2$. Most of the domain (to
the right of the bifurcation point) corresponds to purely real
instability growth rate $\sigma > 0$ with two branches describing
the fast and slow perturbation modes (shown by the solid lines). The
instability growth rate of the fast mode increases with the scaled
activation energy without any limit. The analytical model, Eq.
(\ref{eq31}), predicts the asymptotic increase of the growth rate
for the fast mode as $\sigma \propto (E_{a} / T_{f} )^{2}$ for
$E_{a} / T_{f} \to \infty $. The growth rate of the slow mode
decreases with the scaled activation energy. For this reason it is
expected that the slow mode plays a noticeable role only close to
the bifurcation point, where the growth rates of the fast and slow
modes are comparable. In a small part of the instability domain, for
intermediate values of the scaled activation energy in between the
stability limit and the bifurcation point, $28.2 < E_{a} / T_{f} <
33.2$, the instability growth rate is complex with the real part,
$\sigma > 0$ (dashed line), and imaginary part, $\omega $ (dotted
line). Although this part of the instability domain is rather small,
it indicates the physical outcome of the instability at the
nonlinear stage. Because of the non-zero frequency, $\omega \ne 0$,
it is natural to expect that the instability leads to a pulsating
regime of magnetic deflagration at the nonlinear stage. The
analytical solution, Eq. (\ref{eq31}), obtained within the model of
a thin transition zone (dashed-dotted lines) shows qualitatively the
same stability properties of magnetic deflagration as the numerical
solution. Still, we observe minor quantitative difference between
the analytical model and the numerical solution. According to the
analytical model, the stability limit and the bifurcation point are
expected at $E_{a} / T_{f} = 34.0$ and $E_{a} / T_{f} = 46.8$,
 which differ by approximately 20-30\% from the
respective numerical values. The limited accuracy of the analytical
model is due to the simplifying approximations of 1) a constant
coefficient of energy diffusion and 2) an infinitely thin zone of
energy release. In particular, the shortcomings of the discontinuity
model have been discussed in [\onlinecite{Bychkov-Liberman-94,
Bychkov-review}] in the context of solid propellant combustion.

The numerical solution to the stability problem indicates the
experimental parameters required to observe the unstable
non-stationary regime of magnetic deflagration. We plot the
stability diagram in Fig. 5 in $f - H_{z}$ coordinates using Eq.
(\ref{eq17}); the shading represents absolute value of the
instability growth rate.
\begin{figure}
\includegraphics[width=3.4in,height=2.8in]{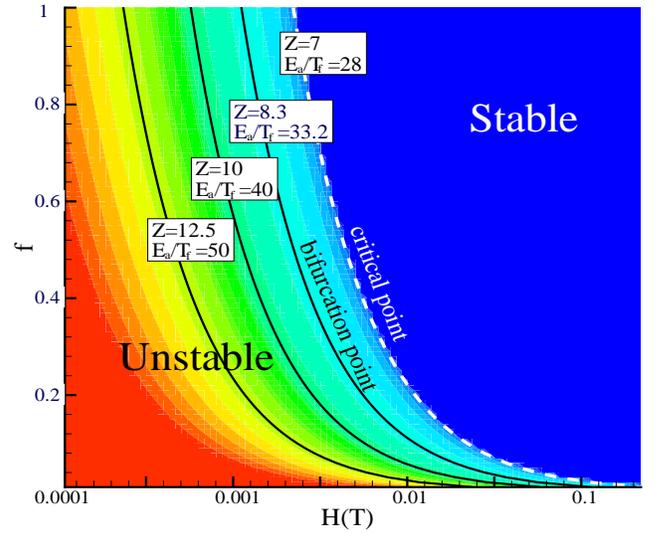}
\caption{ The stability limit (dashed line) of the magnetic
deflagration in coordinates of the dilution factor $f$ and the
magnetic field $H$. The shading shows the instability growth rate.
Other solid lines correspond to the respective constant values of
the Zeldovich number.}
\end{figure}
The dashed white line in Fig. 5 corresponds to the critical value of
the Zeldovich number (the stability limit) obtained numerically.
Magnetic deflagration propagates stationary in the parameter domain
to the right of the stability limit, in the region of a high
magnetic field. To the left of the critical curve, for a low
magnetic field, the stationary magnetic deflagration is unstable,
and we expect a pulsating regime of the deflagration front.
Particularly, in the case of $\rm{Mn}_{12} $ with the dilution level
$f=1/3$ one should expect the instability for the magnetic fields
below $10^{-2}$T, which is possible to achieve experimentally.

In order to understand front dynamics at the nonlinear stage of the
instability, we perform direct numerical simulation of Eqs.
(\ref{eq33})-(\ref{eq34}). In our simulations we use the method of
finite difference for the spatial derivatives and the common fourth
order Runge-Kutta method for the time step integration. Since the
front structure is 1D, then we were able to obtain fine front
resolution together within acceptable time of computational runs.
The results of our simulations, i.e. evolution of the magnetic
deflagration speed, are shown in Fig. 6 for different values of the
scaled activation energy. Figures 6 a, b demonstrate front behavior
close to the stability limit in the stable (a) and unstable (b)
parameter domain. In the first case with $E_{a} / T_{f} = 28.55$,
the velocity perturbation oscillates and decays in time, which
implies a negative instability growth rate. On the contrary, in Fig.
6b with $E_{a} / T_{f} = 28.8$, the amplitude of velocity
oscillations grows in time, which corresponds to the instability
onset. The markers in Fig. 4 show the oscillation frequency of the
magnetic deflagration obtained in direct numerical simulations with
empty circles and filled triangles standing for the stable and
unstable regimes, respectively. As we can see, the direct numerical
simulations are in a good agreement with the solution to the
eigenvalue problem, which concerns both the stability limits and the
oscillation frequency. Still, the instability is rather weak in Fig.
6 b with the oscillations described well by the sine function. The
nonlinear effects become noticeable for higher values of the scaled
activation energy with sharp peaks and smooth troughs in the
oscillations as presented in Fig. 6c for $E_{a} / T_{f} = 30.2$.
\begin{figure}
\includegraphics[width=3.4in,height=1.7in]{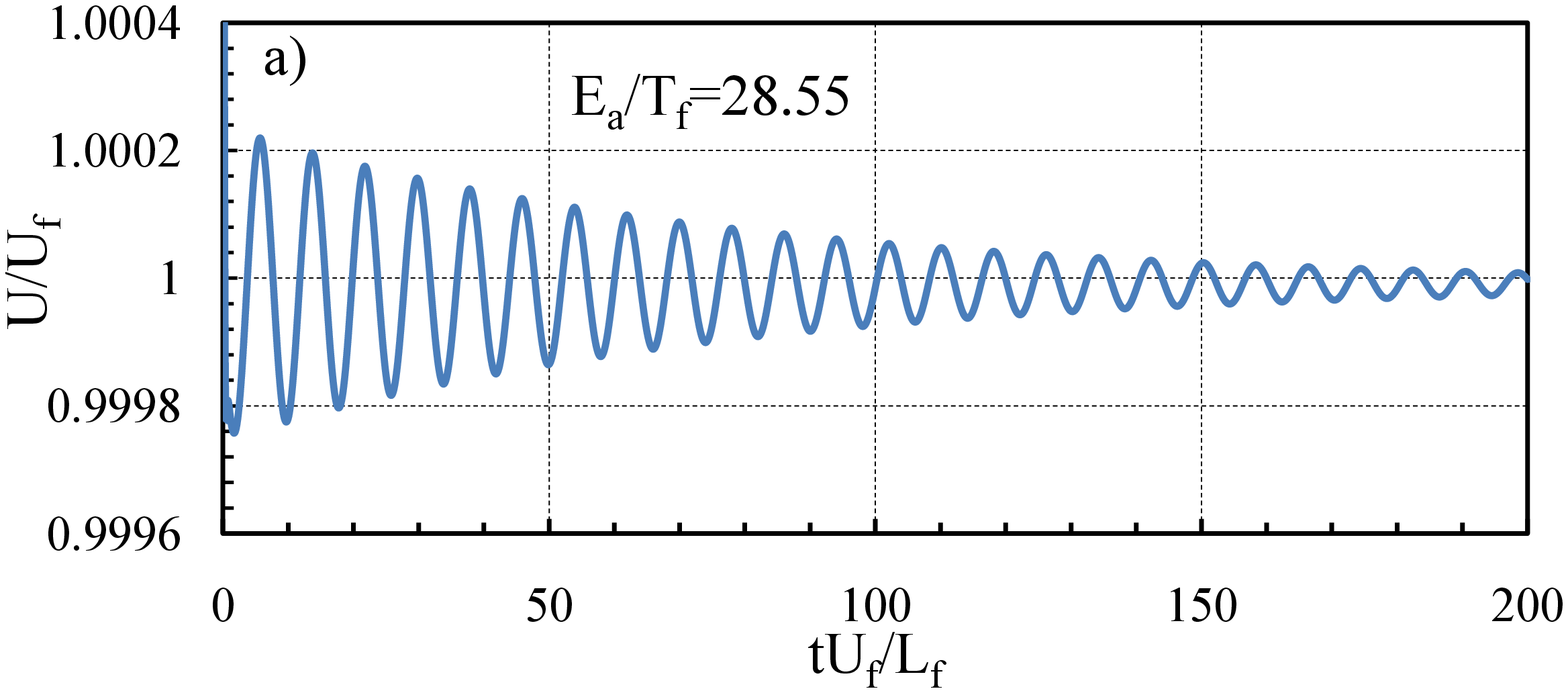}
\includegraphics[width=3.4in,height=1.7in]{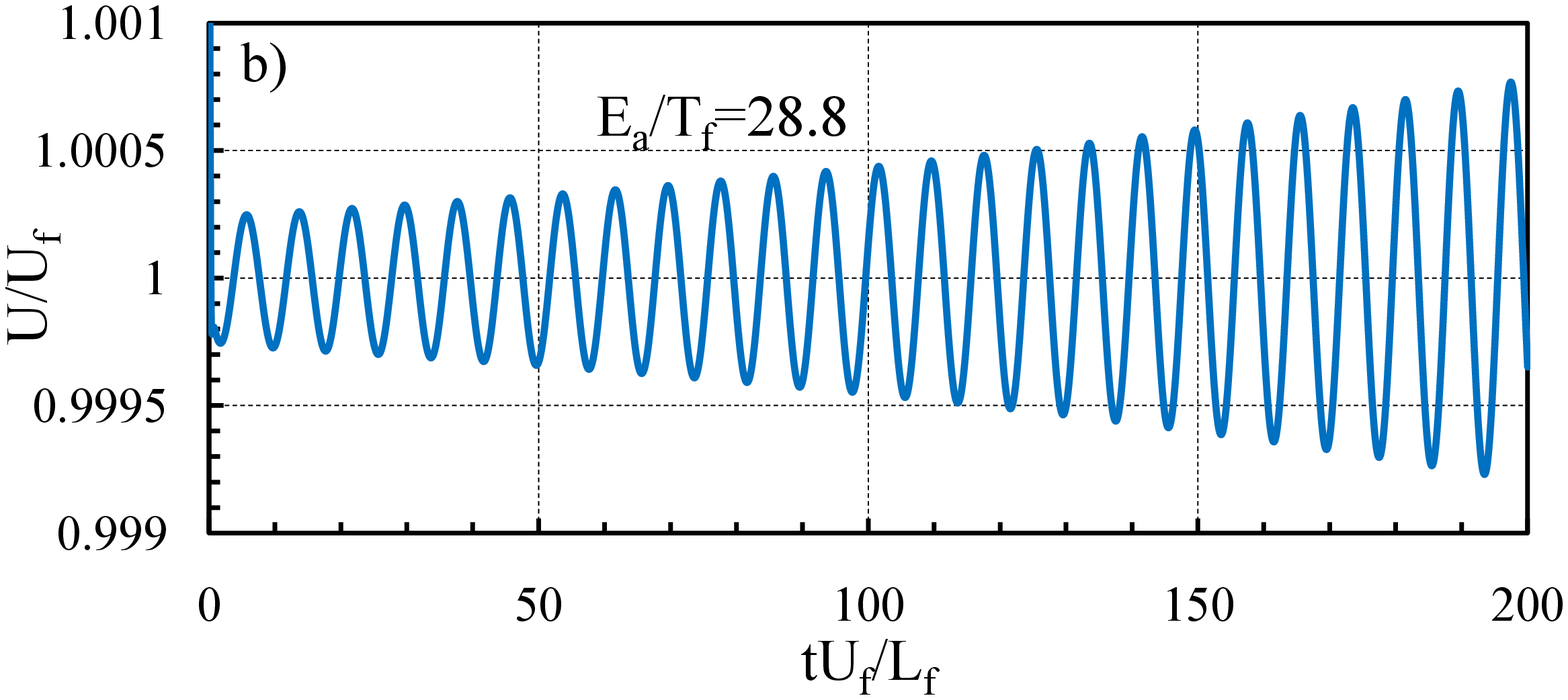}
\includegraphics[width=3.4in,height=1.7in]{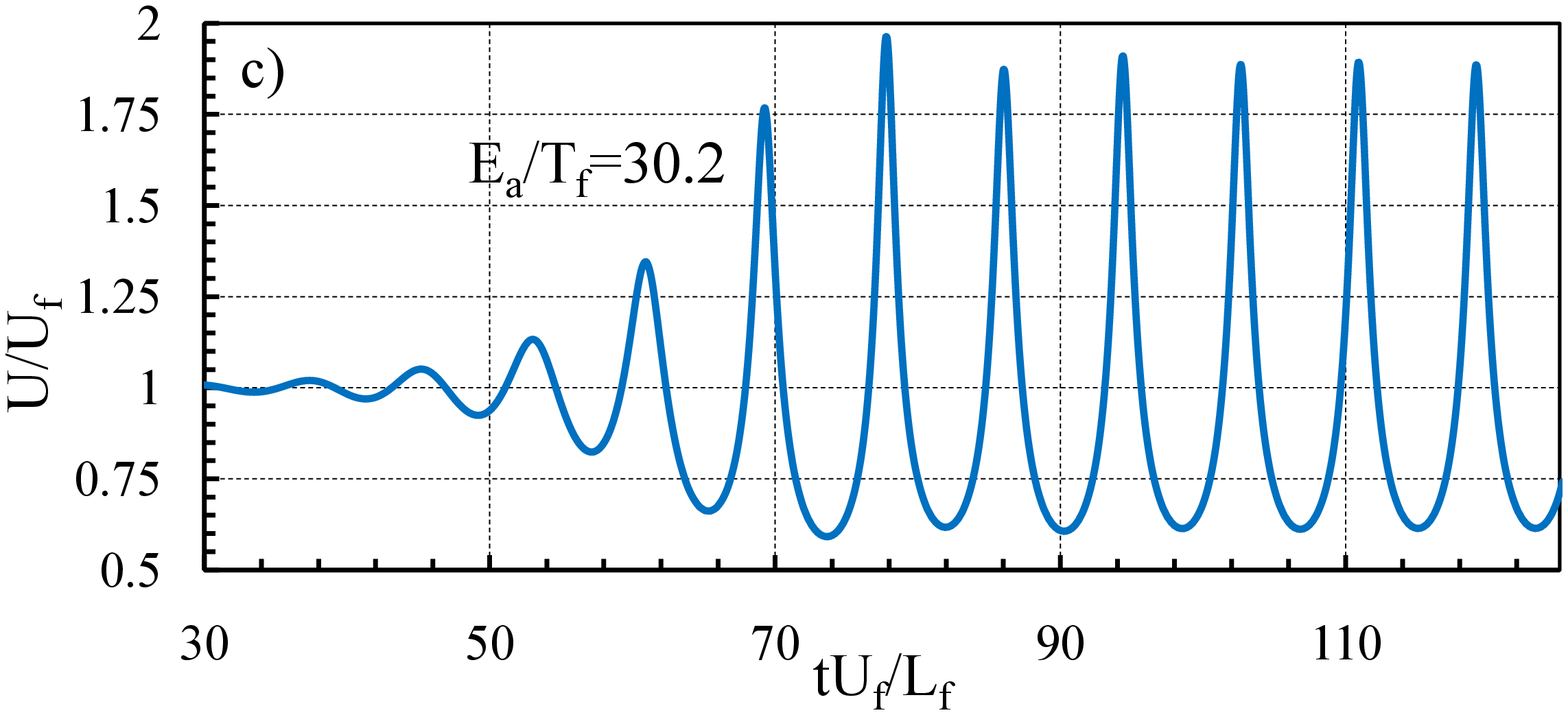}
\includegraphics[width=3.4in,height=1.7in]{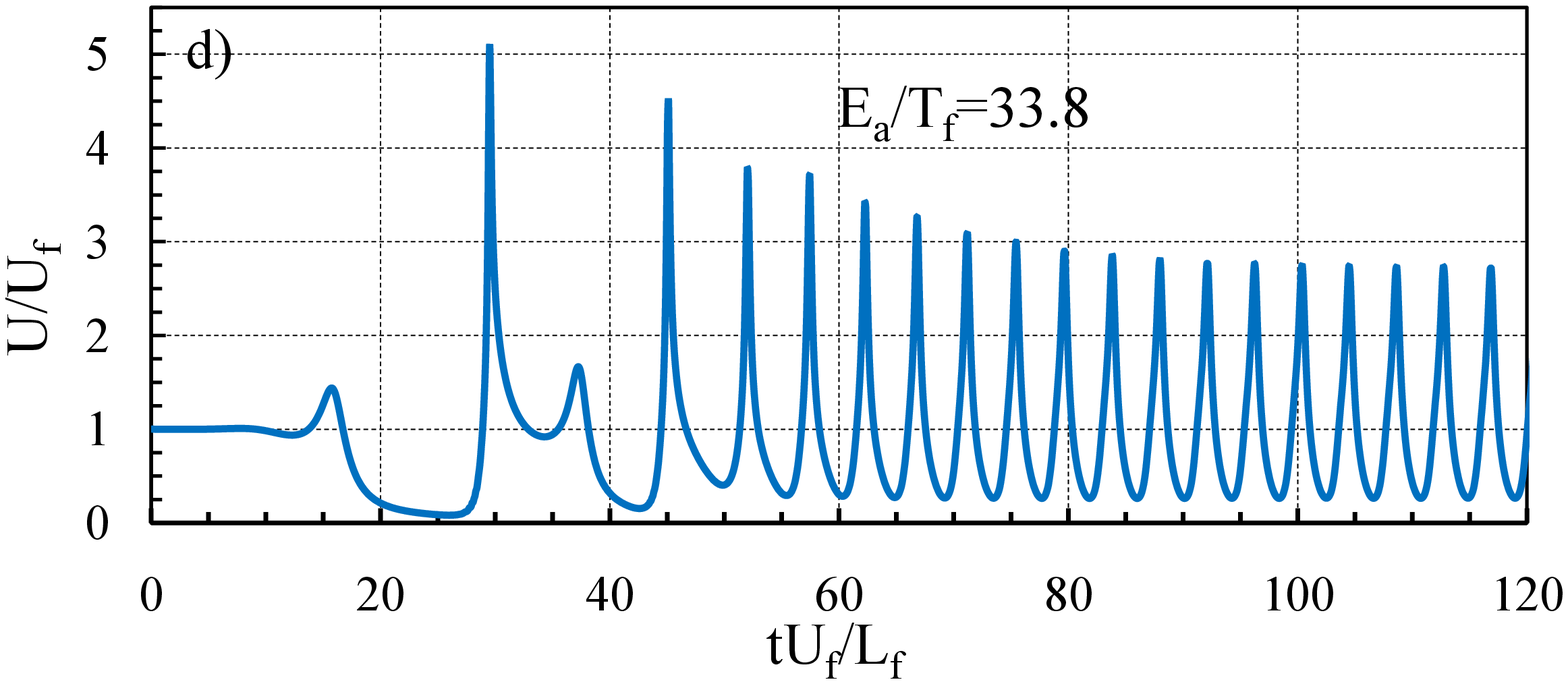}
\includegraphics[width=3.4in,height=1.7in]{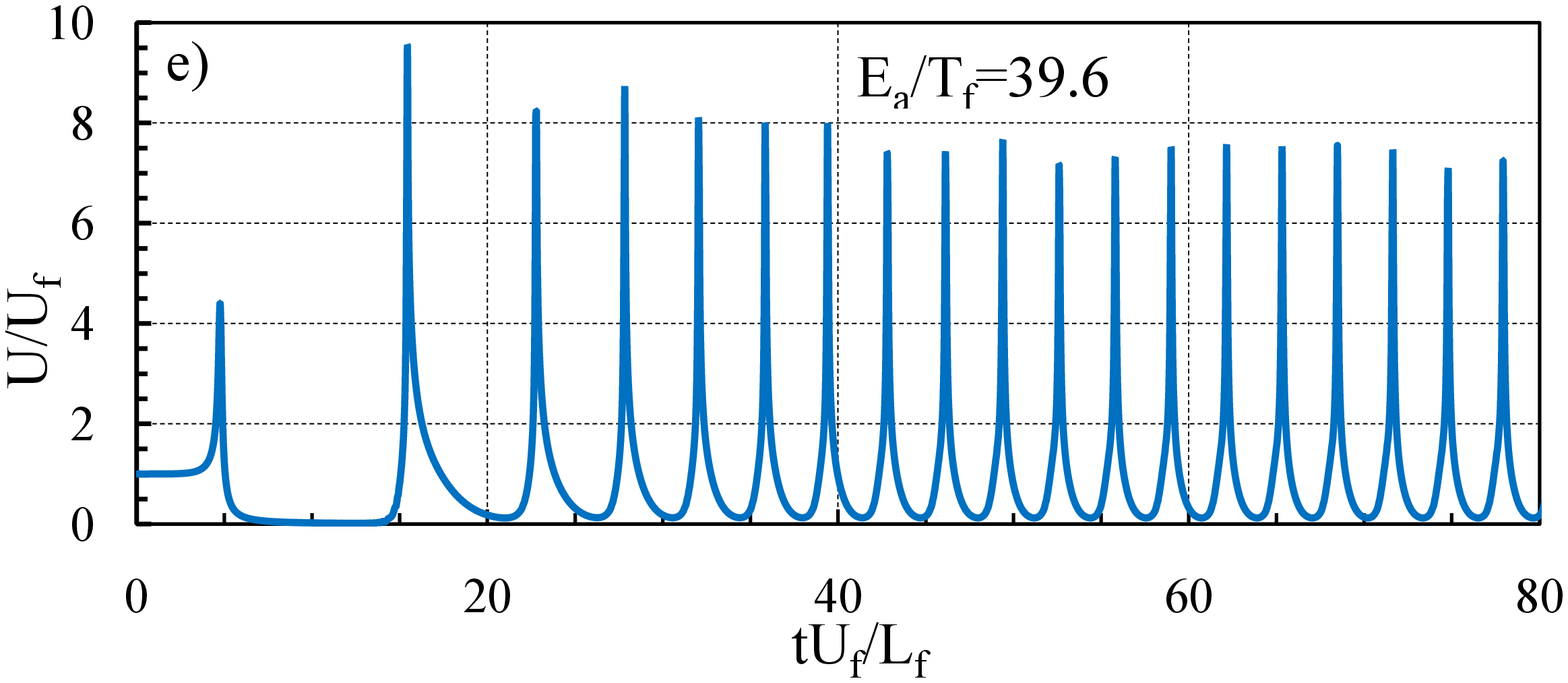}
\caption{Deflagration speed versus time for different values of the
scaled activation energy $E_{a} / T_{f} =
28.55;\;28.8;\;30.2;\;33.8;\;39.6$ for plots (a) - (e),
respectively.}
\end{figure}
The simulations explain the physical mechanism of the obtained instability.
Magnetic deflagration propagates due to two effects: release of the Zeeman
energy in a thin transition zone and transfer of the energy to the cold
layers (preheating) due to thermal conduction. In a stationary regime of
magnetic deflagration, these two processes work at the same rate and balance
each other. However, at high values of the activation energy, the rate of
energy release is too sensitive to temperature in the transition zone. Small
temperature perturbations may increase the ``burning'' rate considerably,
which makes the transition zone sweep fast over the preheated matter until
it comes to cold regions and stops waiting for a new portion of cold matter
to be preheated. As soon as it happens, next pulsation of the front takes
place.

It is interesting that the three figures 6 a, b and c demonstrate
three qualitatively different regimes of magnetic deflagration,
though the activation energy changes only within 5\% from plot (a)
to plot (c). In Fig 6 d we take the activation energy $E_{a} / T_{f}
= 33.8$ close to the bifurcation point and see a complicated front
behavior at the onset of the velocity pulsations. Presumably, the
complicated behavior happens because of two unstable modes with
close instability growth rates at the vicinity of the bifurcation
point. Still, after an initial transition period, the front
pulsations resemble those of Fig. 6 c. Finally, taking the
activation energy noticeably larger than the bifurcation point,
$E_{a} / T_{f} = 40$, we observe even more pronounced nonlinear
features in the front oscillations with even sharper peaks of large
amplitude, shown in Fig. 6e.
The obtained results are qualitatively similar to the pulsation
instability of solid propellant combustion studied in, e.g., 
[\onlinecite{Shkadinsky-71, Frankel-94}].

As we have shown, the scaled activation energy (or the Zeldovich number) is
the main parameter, which controls 1D stability properties of magnetic
deflagration. Still, the problem involves other parameters as well, namely,
power exponents $\alpha $ and $\beta $ in the heat capacity and thermal
conduction, respectively, and the scaled initial temperature of the system $\theta
_{0} $. Figure 7 shows the instability growth rate for different types of
thermal conduction (different values of the power exponent $\beta $, $13 / 3
> \beta > 0)$.
\begin{figure}
\includegraphics[width=3.4in,height=2.6in]{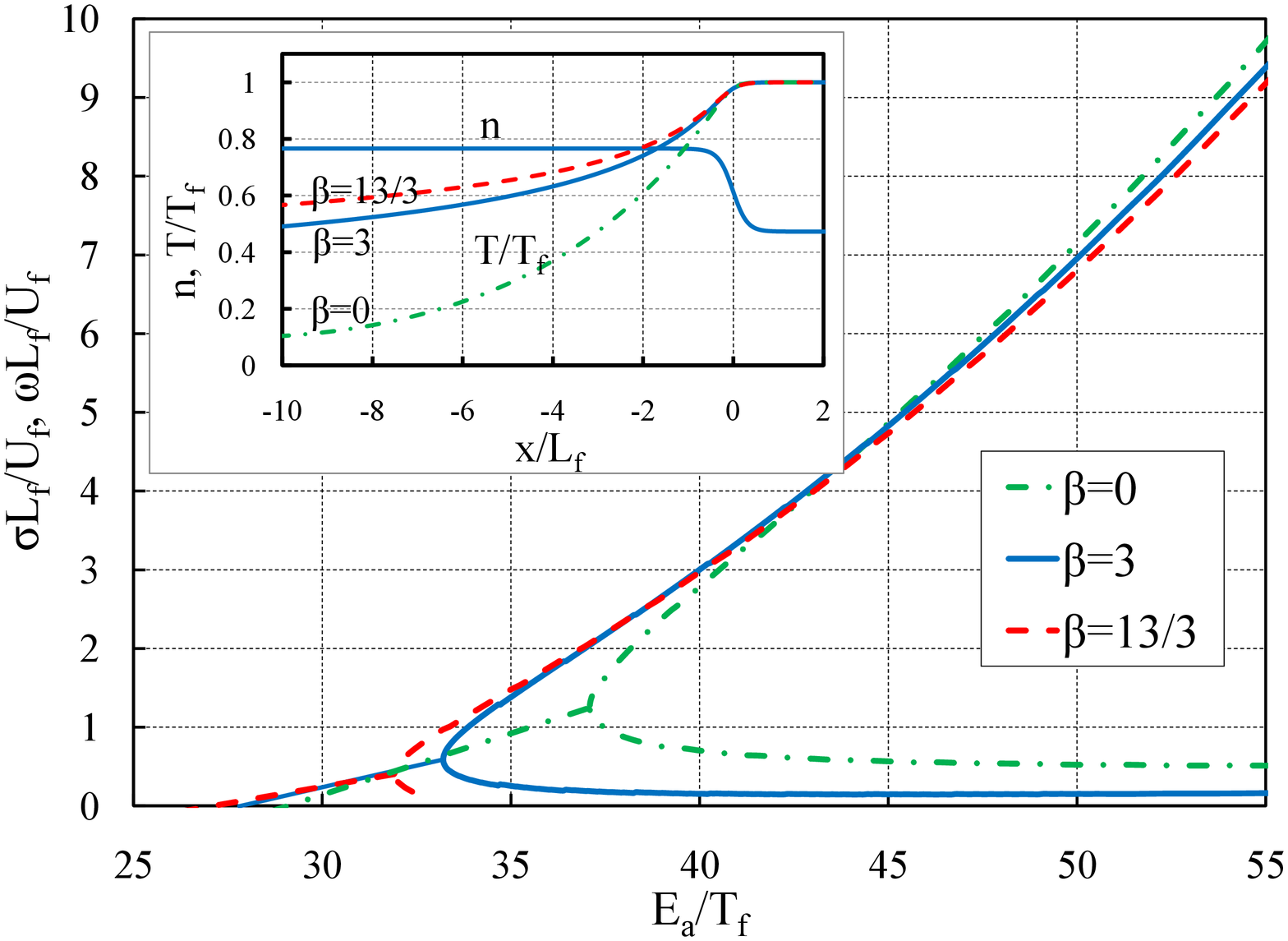}
\caption{The instability growth rate $\sigma $ versus the scaled
activation energy calculated for different values of the power
exponent $\beta $. Other parameters are $\alpha = 3$, $\theta _{0} =
0.2$. The inset presents the respective temperature profiles for
$E_{a} / T_{f} = 30$ and $\beta = 0;\;3$.}
\end{figure}
As we can see, a particular type of the thermal conduction
influences the temperature profile in the stationary magnetic
deflagration noticeably in agreement with Eq. (\ref{eq34}), see the
insert of Fig. 7. At the same time, the stationary profiles of
concentration and energy release remain the same for different
values of the $\beta $-factor. The power exponent $\beta $
determines the total length scale of the deflagration front in the
cold region. However, even considerable variations of $\beta $
modify the instability growth rate only slightly: the stability
limit changes only by 5\% from $E_{a} / T_{f} = 27.9$ to $E_{a} /
T_{f} = 29.4$; the bifurcation point changes within 15\% from $E_{a}
/ T_{f} = 32$ to $E_{a} / T_{f} = 37$. This result demonstrates that
the perturbation modes are strongly localized around the zone of
energy release within the length scale about $L_{f} $ as we
explained in Sec. III. For this reason, considerable modifications of the temperature profile in the far zone on the length scales much larger than $L_{f} $ due to variations of $\beta $ produce a minor effect on the stability properties.
Influence of the initial temperature on the
instability development is even weaker. In particular, changing the scaled initial
temperature from $\theta_{0}=0.2$ to $\theta_{0} = 0.5$, we find
modifications of the stability properties by about 5\%.

Unlike $\beta $ and $\theta_{0} $, the system dimension $\alpha $
(and the power exponent in phonon heat capacity) influences the
critical values of the scaled activation energy $E_{a}/T_{f}$ considerably since it is involved in the
Zeldovich number according to Eq. (\ref{eq15}). At the same time, the critical Zeldovich number changes
slightly for different values of  $\alpha $. For example, we obtain the scaled activation number
$E_{a}/T_{f}=13.7$ and the critical Zeldovich number $Z=6.8$ for $\alpha=1$, which may be compared to  $E_{a}/T_{f}=28.2$ and $Z=7.0$ found for $\alpha =
\;3$. Besides, one may also expect different types of heat
capacity apart from the phonon type, which may modify  the power law Eq. (\ref{eq8}) and the respective deflagration stability properties
\cite{Forminaya-97,Gomes-98,Forminaya-99}.

\section{Summary}

In this paper we have obtained 1D instability of magnetic deflagration in a
medium of molecular magnets. The main parameter of the problem is the
Zeldovich number, which represents the activation energy of the system (in
temperature units) scaled by the temperature at the deflagration front with
a numerical factor depending on the type of heat capacity. We have
demonstrated that the deflagration front becomes unstable when the Zeldovich
number exceeds a certain critical value $Z \approx 7$. We have obtained the
analytical scaling for the instability parameters at the linear stage within
the model of an infinitely thin zone of Zeeman energy release. We have also
solved the eigenvalue stability problem numerically taking into account the
internal structure of the deflagration front. The numerical solution
determines the experimental parameters necessary to observe the instability.
Besides, we have performed direct numerical simulations, which demonstrated
that the instability leads to a pulsating regime of magnetic deflagration at
the nonlinear stage.

\acknowledgments

This work was supported by the Swedish Research Council and by the Kempe
Foundation.

\end{document}